# Biomedical Signals Reconstruction Under the Compressive Sensing Approach

Vesna Mandić, Ivan Martinović
University of Montenegro
Faculty of Electrical Engineering
Džordža Vašingtona bb, 81000, Podgorica, Montenegro
e-mail address: vjesnamandic@gmail.com, ivan.ivo.martinovic@gmail.com

*Abstract*— **The paper analyses the possibility to recover different biomedical signals if limited number of samples is available. Having in mind that monitoring of health condition is done by measuring and observing key parameters such as heart activity through electrocardiogram or anatomy and body processes through magnetic resonance imaging, it is important to keep the quality of the reconstructed signal as better as possible. To recover the signal from limited set of available coefficients, the Compressive Sensing approach and optimization algorithms are used. The theory is verified by the experimental results.**

## I. INTRODUCTION

Advances in technology ensues tendencies for widening of technological methods to all aspects of human life. Primary category is, and rightfully so, health. Medicine and technology are becoming more dependent on each other, and thus strongly bonded together, with the underlying reason that medicine without technology cannot be optimally efficient, while technology without medicine is not sufficiently impelled toward progress. Every clinic is obligated to provide primary health protection to its patients, a quality which is reflected, among other standards, in a capability of providing effective and efficient methods of diagnostics.

Monitoring and checking of health conditions of potential patients is conducted by measuring blood pressure, recording electric activity of heart by electrocardiograph (ECG), respiratory examination, magnetic resonance imaging etc. All these methods require reliable instruments, which are based on verified scientific principles.

Electrocardiogram [1]-[7] is a graphical waveform whose spikes and drops mimics a heart function. It is obtained by placing electrodes on strategic points on patient's body. In this way, a signal comprised of following curves is obtained: contraction of heart atrium (P-complex), and contraction of heart vestibules (QRS- and T- complexes). Any deviation from normal shape of this signal indicates to a possible heart disorder. Typical computerized system for digital signals processing requires a large amount of data which is difficult to transfer and store. The main goal of any compression is a reduction of the amount of data, with guaranteed preservation of its important characteristics.

The Compressive Sensing (CS) [8]-[25] raised as an approach that can provide complete information of the signal if there is a small set of signal coefficients available. Therefore, this approach founds many applications, and one of them is in the biomedical signals reconstruction. It is important to emphasize that sufficient number of samples for signal recovering in CS scenario is smaller than the one specified in Nyquist-Shannon theorem.

In this paper, several types of biomedical signals are observed and the possibility for CS reconstruction using different number of available samples is tested. The reconstruction is done by using the CS approach and optimization algorithms based on the total variation (TV) minimization. The errors between original and reconstructed signals are measured, in order to test what is the minimal number of samples required to successfully recover the signal of interest.

## II. BIOMEDICAL SIGNALS PROCESSING

a) Biomedical signals are the outcome of the observation of physiological activities of organisms. Observations include the whole body and they are conducted using different instruments. Signals obtained through these observations generally carry a large amount of data. Efficient compression of these signals is important for storing data and transmission over the digital telecommunication network and telephone line.

In this paper, three different types of biomedical signals, under the CS scenario, are observed: ECG, sphygmomanometer signal and respiration monitor's signal.

ECG is, in fact, observing the electric current generated by the heart which is conducted through the wires or transmitted wireless by radio to the recording device. The device consists of an amplifier that magnifies the electric signals and galvanometer with inscription needle. The needle inscribes a positive or negative deflection which mimics heart rhythm. Through this received signal one can measure heart rate, know the heart rhythm, calculate electrical axis of the heart and analyze sequentially the different waves and segments of the ECG [2].

Sphygmomanometer is used for measuring blood pressure. Blood pressure is the amount of pressure exerting on the walls of blood vessels within the body, measured in mmHg. It involves calculating the pressure during the contraction (systole) and relaxation (diastole) of the heart. Monitoring of blood pressure is important for controlling hypertension or high blood pressure. Certain groups of people merit special consideration for the measurement of blood pressure—because of age, body habitus, life situation [2] etc.

Respiration monitor with its measuring element follows depth and frequency of breathing. It is done by strapping a belt with measuring elements around the chest of the patient [2].

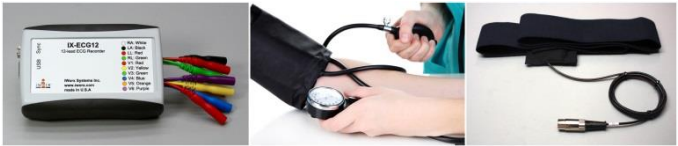

Figure 1: Instruments: ECG, Sphygmomanometer and Respiration monitor

b) Until recently, signal acquisition is made according to the Sampling theorem – with a sampling frequency at least two times maximal signal frequency. After the acquisition, compression is a common step in the majority of the applications. CS aims at performing the compression during the acquisition. It is achieved by acquiring sparse signals according to the CS rules – random sampling showed to be the common CS sampling approach. Various algorithms that recover signal information from a limited number of signal samples are developed. Some of them are more complex but also more precise, such as convex optimization [8]-[12], but there are also computationally less demanding approaches, such as greedy approaches [12],[18], threshold-based solutions [12],[14],[16],[20], etc. The CS-based reconstruction procedure is described in Section III.

### III. CS BASED RECONSTRUCTION PROCEDURE

An *N*-dimensional signal could be written in terms of its transform domain representation, as:
$$\mathbf{h} = \sum_{i=1}^{N} H_i \theta_i = \mathbf{\Theta H}, \quad (1)$$

where $H_i$ is weighting coefficient, $\theta_i$ is basis vector, **Θ** denotes $N \times N$ transform matrix. Domain **Θ** is domain of sparsity. If we model random acquisition of the *M* signal measurements (where $M \ll N$ holds) with the measurement matrix **ϒ** (of size $M \times N$), then the vector of the available signal samples $\mathbf{h}_a$, of $M \times 1$ size, is modeled as:
$$\mathbf{h}_a = \mathbf{\Upsilon h} = \mathbf{\Upsilon \Theta H} = \mathbf{\Delta H}, \quad (2)$$
where **Δ** is a CS matrix.

Having in mind that biomedical signals we observed in this paper, are not sparse neither in time nor in the frequency domain, the commonly applied 1D reconstruction algorithms failed to provide an accurate signal recovery (even if we use a large number of measurements). Therefore, the reconstruction in 2D domain is proposed. It is done by transforming 1D signal to 2D version, undersampling and then applying the TV minimization. TV minimization is usual method for the reconstruction of the under-sampled 2D data, and it is based on the image gradient minimization [21]-[25].

Firstly, 1D signal **h** is reshaped into the matrix **I** that will act as an image to be under-sampled and recovered. The reshaping is done column-wise and quadratic matrix is obtained [22]:
$$\mathbf{I} = \rho(\mathbf{h}, \sqrt{N}, \sqrt{N}), \quad (3)$$

where ρ denotes the vector-matrix conversion operator. The image is of $\sqrt{N} \times \sqrt{N}$ size. In the case when $\sqrt{N}$ is not an integer, the signal is zero-padded in order to obtain an integer value for $\sqrt{N}$. As a domain of sparsity, a two dimensional discrete cosine transform (2D DCT) domain is considered.

Let us now denote a set of image measurements as $\mathbf{P}_a$, taken from the 2D DCT domain, in a random manner from the zig-zag reordered 2D DCT coefficients (the matrix **Θ** corresponds to the 2D DCT matrix). The image is reconstructed from the acquired measurements by solving an optimization problem, defined as:
$$\min_{\mathbf{H}} \zeta(\mathbf{H}) \text{ subject to } \mathbf{h}_a = \mathbf{\Delta F}, \quad (4)$$

where **ζ** denotes the TV operator, defined as a sum of the magnitudes of discrete gradient at each point:
$$\zeta(\mathbf{H}) = \sum_{i,j} \|p_{i,j}\|_2 \quad (5)$$

where the gradient approximation for the pixel *ij*, $p_{i,j}$, is described as [22]:
$$p_{i,j}\mathbf{H} = \begin{bmatrix} H(i+1,j) - & H(i,j) \\ H(i,j+1) - & H(i,j) \end{bmatrix}, \quad (6)$$

### IV. EXPERIMENTAL RESULTS

*Example 1: ECG signal*

In the first example, a part of ECG signal is observed, and its reconstruction using relatively small number of samples is considered. The 1D signal was converted into the image, according to equation (3). Different number of measurements are taken, and reconstruction quality is measured by calculating mean square error (MSE) between original and reconstructed 1D signal. Mean value of the signal is equal to -23.73. After image reconstruction, the 1D signal is extracted using procedure which is reverse to image forming.

The original and reconstructed images, as well as original and reconstructed signals are shown in Figure 2. Signal is reconstructed using 40% of the coefficients. MSE error of the reconstruction, for different number of available samples, is shown in Table 1.

*Example 2: Sphygmomanometer signal*

In the second example, part of the sphygmomanometer signal was observed. As in the previous example, the 1D signal is first converted to the image, and then certain percent of the samples is chosen randomly. The reconstruction was done using the acquired samples by applying the TV minimization. The 1D signal was extracted.

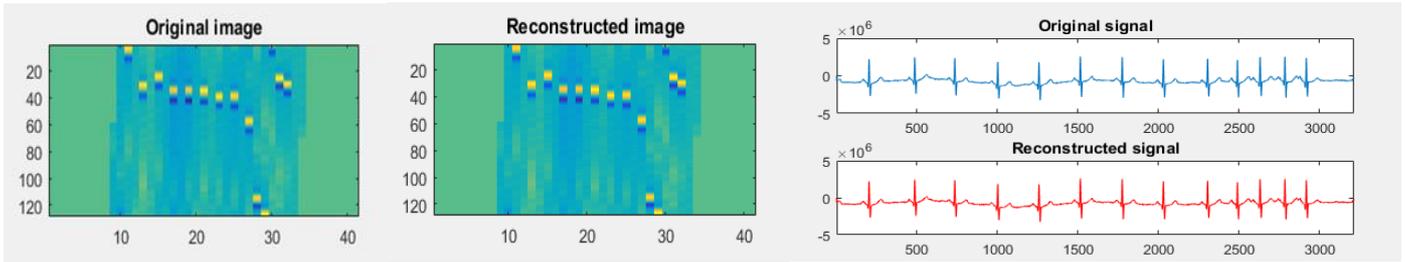

Figure 2.ECG signal. First column: Original image; Second column: Reconstructed image; Third column: Original 1D signal (blue) and reconstructed 1D signal (red). Signal is reconstructed by using 40% of the available samples

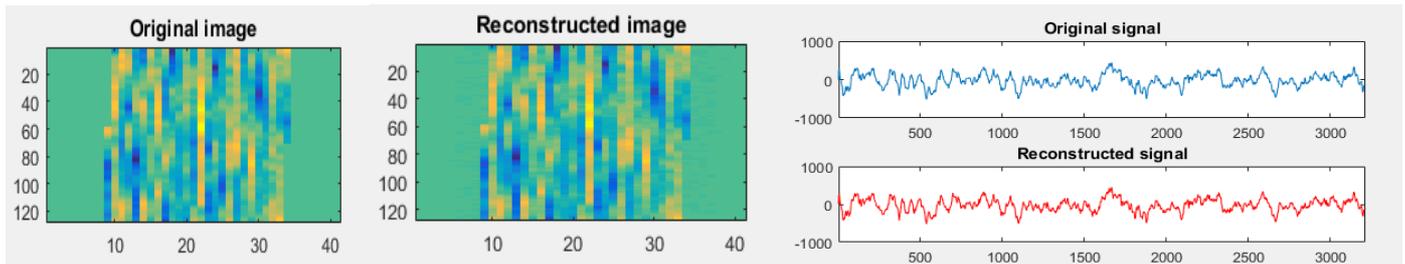

Figure 3.Sphygmomanometer signal. First column: Original image; Second column: Reconstructed image; Third column: Original 1D signal (blue) and reconstructed 1D signal (red). Signal is reconstructed by using 45% of the available samples

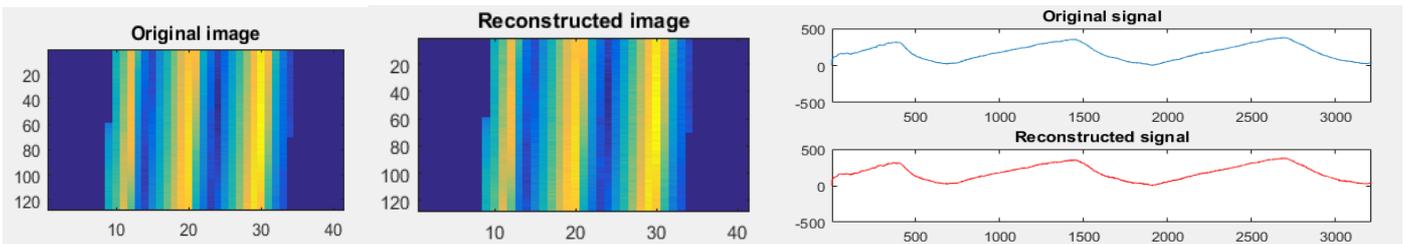

Figure 4.Respiration Monitor signal. First column: Original image; Second column: Reconstructed image; Third column: Original 1D signal (blue) and reconstructed 1D signal (red). Signal is reconstructed by using 45% of the available samples

TABLE I. MSEs of original and reconstructed signals for different percentage of available samples

| Percentage of the available samples | Signal from example 1 | Signal from example 2 | Signal from example 3 | Percentage of the available samples | Signal from example 1 | Signal from example 2 | Signal from example 3 |
|---|---|---|---|---|---|---|---|
| 30% | 52.089 | 373.27 | 14.43 | 65% | 1.2356 | 1.242 | 0.10072 |
| 35% | 28.819 | 176.51 | 7.6239 | 70% | 0.68938 | 0.028419 | 0.018274 |
| 40% | 21.793 | 72.31 | 3.8149 | 75% | 0.27915 | 0.018064 | 0.016685 |
| 45% | 13.55 | 31.305 | 1.6361 | 80% | 0.18973 | 0.012232 | 0.014737 |
| 50% | 7.5845 | 11.525 | 0.61697 | 85% | 0.040406 | 0.0092771 | 0.012061 |
| 55% | 4.033 | 3.418 | 0.34745 | 90% | 0.019434 | 0.0076873 | 0.010572 |
| 60% | 2.6299 | 1.5991 | 0.091887 | | | | |

The original and reconstructed images, as well as original and reconstructed 1D signals are shown in Figure 3. The 1D and 2D signals shown in Figure 3 are reconstructed by using 40% of the measurements. Mean value of the signal is equal to -4.87.

*Example 3: Respiration Monitor signal*

In the third example, part of the respiration monitor signal is considered. As in the previous cases, the signal is converted to 2D form and undersampled. The image reconstruction using the acquired samples is done with TV minimization. The original and reconstructed signals are shown in Figure 4, while MSEs for different number of available samples are shown in Table 1. Mean value of the signal is equal to 35.38. The signals are reconstructed by using 45% of the available samples.

Based on the experimental result it is shown that the observed biomedical signals can be successfully recovered if less then 50% of the samples is available. This reduction of the number of samples, required for successful signal analysis and processing, can be of great interest in biomedicine, to increase the processing speed. Also, having in mind that information about patients are constantly recorded and stored, this approach can help in decreasing the storage requirements as well.

## V. CONSLUSION

Application of CS approach on biomedical signals is considered in this paper. The observed are ECG signal, blood pressure and respiration monitor signal. Signals are undersampled and reconstructed by using the CS approach. Different number of available samples are used in all considered cases. The reconstruction quality is measured visually and by calculating the MSE. It is shown that the signals can be reconstructed if less than 50% of the total number of samples is available. Samples are chosen randomly and 2D reconstruction method is applied. This research showed that compressive sensing method may have benefits in reducing time spent in the process of monitoring health condition and decrease amount of data for storing and transferring data through communication systems.